\documentclass[10pt,twocolumn,reqno,amsmath,amssymb,aps,prl,superscriptaddress,floatfix]{revtex4-2}
\usepackage{lmodern}            
\usepackage{graphicx} 
\usepackage{enumitem}
\usepackage{amsmath,amssymb,amsfonts,mathtools}
\usepackage{bm}                 
\usepackage{multirow}
\usepackage[version=4]{mhchem}
\usepackage[caption=false]{subfig} 
\usepackage{xcolor}
\usepackage{booktabs}
\usepackage{placeins}
\usepackage{siunitx}            
\usepackage{romannum}

\usepackage[colorlinks=true,linkcolor=blue,citecolor=blue]{hyperref}
\usepackage[all]{hypcap}
\hypersetup{
  colorlinks=true,
  linkcolor=blue,
  citecolor=blue,
  urlcolor=blue,
  hypertexnames=false
}

\begin{document}

\title{Pseudo-superconducting-diode effect in ferroelectric Josephson junctions}
\author{Yaozu Tang}
\affiliation{Kavli Institute of Nanoscience, Delft University of Technology, Lorentzweg 1, 2628 CJ Delft, The Netherlands}
\affiliation{Department of Quantum Nanoscience, Delft University of Technology, Lorentzweg 1, 2628 CJ Delft, The Netherlands}
\author{Nienke ten Haaf}
\affiliation{Department of Quantum Nanoscience, Delft University of Technology, Lorentzweg 1, 2628 CJ Delft, The Netherlands}
\affiliation{Department of Quantum Technology, Netherlands Organisation for Applied Scientific Research (TNO), Stieltjesweg 1, 2628 CK Delft, The Netherlands.}
\author{Artem Bondarenko}
\affiliation{Department of Quantum Nanoscience, Delft University of Technology, Lorentzweg 1, 2628 CJ Delft, The Netherlands}
\author{Mazhar N. Ali}
\affiliation{Kavli Institute of Nanoscience, Delft University of Technology, Lorentzweg 1, 2628 CJ Delft, The Netherlands}
\affiliation{Department of Quantum Nanoscience, Delft University of Technology, Lorentzweg 1, 2628 CJ Delft, The Netherlands}
\author{Yaroslav M. Blanter}
\affiliation{Kavli Institute of Nanoscience, Delft University of Technology, Lorentzweg 1, 2628 CJ Delft, The Netherlands}
\affiliation{Department of Quantum Nanoscience, Delft University of Technology, Lorentzweg 1, 2628 CJ Delft, The Netherlands}
\date{\today}

\begin{abstract}
    The superconducting diode effect (SDE), characterized by unequal critical supercurrents in opposite current directions, enables supercurrent rectification. We propose a magnetic-field-free pseudo-superconducting-diode effect in ferroelectric Josephson junctions with broken inversion symmetry. Using a coupled dynamical model that combines a polarization-dependent RCSJ description with Landau-Khalatnikov-Tani ferroelectric dynamics, we show that ferroelectric polarization switching induces asymmetric critical and retrapping currents under current sweeps. The resulting nonreciprocity is highly tunable via ferroelectric parameters and the sweep protocol and remains robust at finite temperatures. Our work identifies ferroelectric Josephson junctions as a promising platform for magnetic-field-free nonreciprocal superconducting devices.
\end{abstract}

\maketitle

\textit{Introduction.---}Nonreciprocal transport in semiconducting diodes is a central concept in condensed matter physics and underlies modern electronics~\cite{shockley1949,sze2008}. In superconductors, the analogous superconducting diode effect (SDE) has recently attracted significant attention~\cite{ando2020,wu2022,jiang2022,nadeem2023}. It manifests itself as unequal critical currents in opposite current directions, allowing a dissipationless supercurrent to flow preferentially in one direction. SDEs have been reported in both junction-free superconductors~\cite{silaev2014,ando2020,bauriedl2022,narita2022,lin2022,du2024} and Josephson junctions (JDEs)~\cite{baumgartner2022,baumgartner2022a,wu2022,pal2022,golod2022,jeon2022,gupta2023,trahms2023}.

Most SDE mechanisms require broken inversion and time-reversal symmetries~\cite{wakatsuki2017,tokura2018,daido2022,davydova2022,he2022}, often through spin-orbit coupling, magnetochiral anisotropy, magnetic textures, or external magnetic fields~\cite{nadeem2023}. Field-free mechanisms are therefore highly desirable for scalable superconducting circuits. Several routes have been proposed in noncentrosymmetric but time-reversal-symmetric systems, including electron-correlation effects~\cite{morimoto2018}, asymmetric charging energy~\cite{misaki2021}, vortex-antivortex dynamics~\cite{itahashi2022}, higher-harmonic current-phase relations~\cite{souto2022a,fominov2022a,seoanesouto2024}, and asymmetric quasiparticle-induced retrapping~\cite{steiner2023}. However, experimental demonstrations remain limited~\cite{wu2022,liu2024,nagata2025}, and in some cases intrinsic time-reversal-symmetry breaking has been suggested~\cite{qi2025,ma2025a}. This motivates alternative field-free routes to diode-like superconducting transport.

Here we propose such a route in an inversion-asymmetric ferroelectric Josephson junction (FE-JJ), building on recent progress in polarization-controlled supercurrent transport~\cite{tang2026,rahmonov:hal-03295760,suleiman2021,badarne2025}. The effect arises from the combination of a polarization-dependent critical current~\cite{tang2026,badarne2025} and current-induced ferroelectric polarization switching~\cite{badarne2025}. When the bias current is swept beyond the polarization-switching threshold, different critical currents are reached in the two polarization states and therefore become unequal. If the sweep remains below the switching threshold, the polarization state remains unchanged and the critical currents are identical. The resulting nonreciprocity is thus a history-dependent nonequilibrium effect, which we call the \emph{pseudo-superconducting-diode effect}.

We describe coupled phase--polarization dynamics using a polarization-dependent resistively and capacitively shunted junction (RCSJ) model~\cite{stewart1968,mccumber1968,tinkham2004} combined with Landau-Khalatnikov-Tani dynamics~\cite{tani1969,sivasubramanian2004}. Numerical simulations show that polarization switching in the resistive state produces asymmetric critical and retrapping currents. We further demonstrate that the effect is tunable through ferroelectric parameters and the current-sweep protocol and remains robust against thermal fluctuations. Our work identifies FE-JJs as a promising platform for field-free nonreciprocal superconducting devices and highlights the rich physics arising from the interplay of ferroelectricity and superconductivity in nonequilibrium settings \cite{pitton2026,donaire2026}.

\begin{figure}[ht]
    \centering
    \includegraphics[width=0.9\columnwidth]{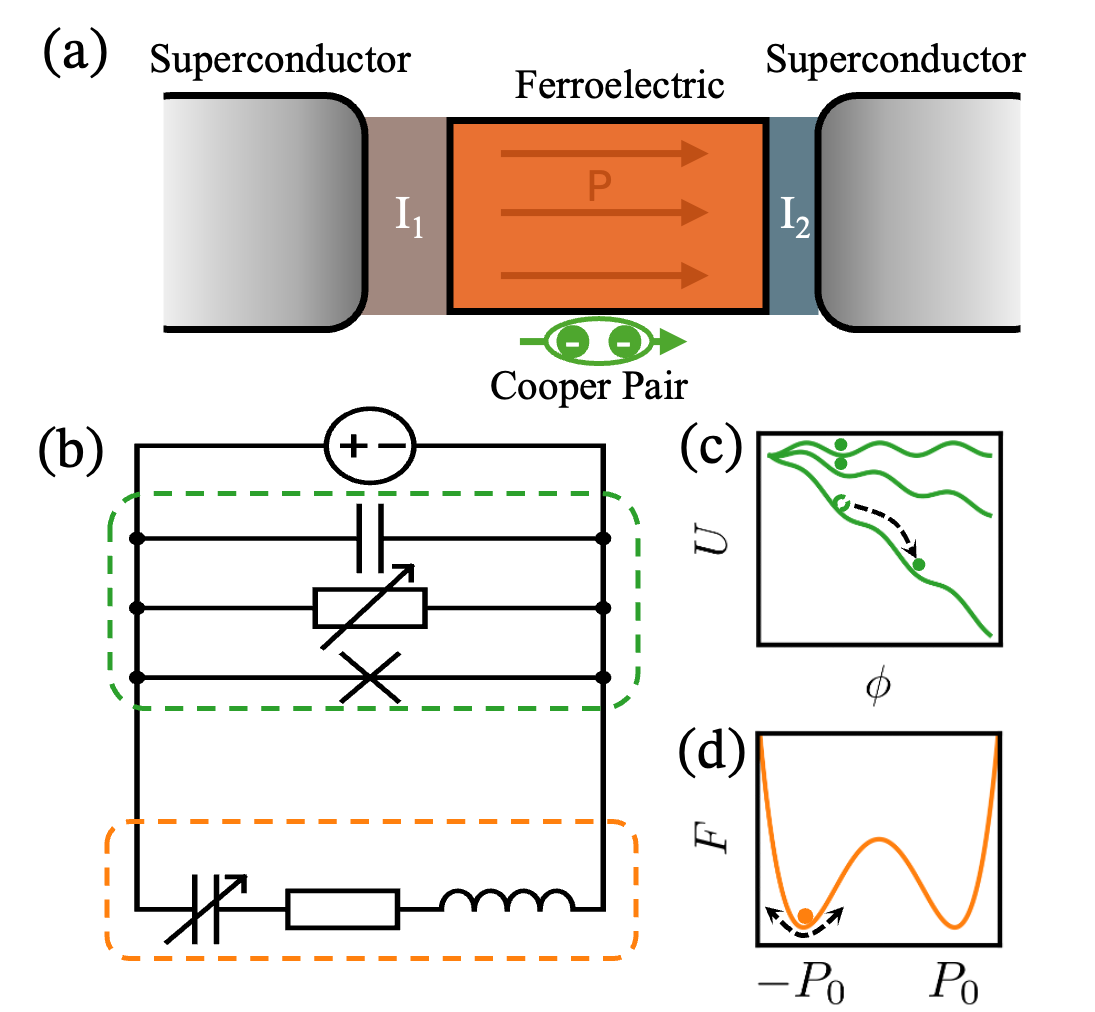}
    \caption{\textbf{Junction and potential landscapes.} 
    (a) Schematic of a FE-JJ with broken inversion symmetry due to interfacial insulating layers (I\(_1\) and I\(_2\)). 
    (b) Equivalent circuit: the green box denotes the conventional RCSJ model, while the orange box represents ferroelectric dynamics. 
    (c) Tilted washboard potential of the RCSJ model. 
    (d) Landau double-well potential of the ferroelectric.}
    \label{fig:system}
\end{figure}


\textit{Model.---}We consider an inversion-asymmetric FE-JJ, as sketched in Fig.~\hyperref[fig:system]{1(a)}. Two superconducting electrodes are separated by a ferroelectric barrier and asymmetric interfacial insulating layers. The dynamics are governed by coupled differential equations for the superconducting phase and the ferroelectric polarization~\cite{rahmonov:hal-03295760,badarne2025}. The Josephson dynamics are described by the RCSJ model, in which the junction is represented by an ideal Josephson element with $I_s = I_c \sin\phi$, a shunt resistance $R_0$, and a capacitance $C_0$ [Fig.~\hyperref[fig:system]{1(b)}]. For a total bias current $I$, the conventional RCSJ equation reads~\cite{stewart1968,mccumber1968,tinkham2004}

\begin{equation}
    I = C_0 \dot{V} + \frac{V}{R_0} + I_c \sin\phi,
    \label{eq:RCSJ}
\end{equation}
with $\phi$ the phase difference and voltage $V=(\hbar/2e)\dot{\phi}$. This equation can be interpreted as the motion of a phase particle in a tilted washboard potential [Fig.~\hyperref[fig:system]{1(c)}].

In an inversion-asymmetric FE-JJ, the ferroelectric polarization modifies the tunneling conductance and hence the critical current~\cite{tang2026}. For small polarization-induced corrections, the resistance and critical current can be linearized in $P$, yielding
\begin{equation}
    I = C_0 \dot{V} + \left(1-\theta P\right)\frac{V}{R_0} 
      + \left(1-\theta P\right) I_c \sin\phi,
    \label{eq:RCSJ-modified}
\end{equation}
where $\theta$ is a device-dependent asymmetry factor~\cite{tang2026}. Equation~\eqref{eq:RCSJ-modified} reduces to the conventional RCSJ equation when the polarization correction vanishes.

The ferroelectric free-energy density is described by the Landau-Ginzburg-Devonshire (LGD) theory,
\begin{equation}
    F = -\frac{\alpha}{2}P^2 + \frac{\beta}{4}P^4 - EP,
    \label{eq:LGD}
\end{equation}
where $E$ is the electric field across a ferroelectric barrier of thickness $d$ and is assumed to be parallel to the polarization. Below the ferroelectric transition temperature, $\alpha>0$ and $\beta>0$, so the field-free double-well potential has two minima at $P=\pm P_0$, where $P_0=\sqrt{\alpha/\beta}$ [Fig.~\hyperref[fig:system]{1(d)}].

The polarization dynamics follow the Landau-Khalatnikov-Tani (LKT) equation~\cite{tani1969,sivasubramanian2004},
\begin{equation}
    m_p \ddot{P} + \gamma_p \dot{P} = -\frac{\partial F}{\partial P},
    \label{eq:LKT}
\end{equation}
where $m_p$ is the polarization inertia and $\gamma_p$ is a phenomenological damping constant. A time-dependent polarization generates a polarization current $I_p=S\dot{P}$, with $S$ the junction area, which couples the ferroelectric dynamics back to the Josephson dynamics. Equivalently, the LKT dynamics can be represented by a shunted circuit consisting of an effective inductor $L_p=m_p d/S$, a resistor $R_p=\gamma_p d/S$, and a nonlinear capacitor $C(P)=S/[d\alpha(P^2/P_0^2-1)]$ [Fig.~\hyperref[fig:system]{1(b)}].

Combining Eqs.~\eqref{eq:RCSJ-modified} and \eqref{eq:LKT} with the Josephson relation and $I_p=S\dot{P}$, we obtain the coupled dimensionless equations at zero temperature,
\begin{align}
    \mathcal{I} &= \beta_c \dot{\mathcal{V}} + (1-\tilde{\theta} \mathcal{P})\mathcal{V} 
        + (1-\tilde{\theta} \mathcal{P})\sin\phi + \mathcal{I}_p, \label{eq:ODEs-RCSJ} \\
    \mathcal{V} &= \mu \ddot{\mathcal{P}} + \gamma \dot{\mathcal{P}} 
        + A(\mathcal{P}^2-1)\mathcal{P} + \frac{\tilde{\theta}}{\nu}\cos\phi, \label{eq:ODEs-LKT} \\
    \mathcal{V} &= \dot{\phi}, \label{eq:ODEs-Josephson} \\
    \mathcal{I}_p &= \nu \dot{\mathcal{P}}. \label{eq:ODEs-Ip}
\end{align}
Here we introduce the following dimensionless variables and parameters: current $\mathcal{I} = I/I_c$, voltage $\mathcal{V} = V/(I_c R_0)$, polarization current $\mathcal{I}_p = I_p/I_c$, reduced polarization $\mathcal{P} = P/P_0$, Stewart--McCumber parameter $\beta_c = (2e/\hbar) I_c R_0^2 C_0$, dimensionless asymmetry factor $\tilde{\theta}=\theta P_0$, polarization inertia $\mu = (2e/\hbar)^2 m_p I_c R_0 P_0 d$, Landau coefficient $A = \alpha P_0 d/(I_c R_0)$, polarization damping $\gamma = (2e/\hbar) \gamma_p P_0 d$, and $\nu = (2e/\hbar) R_0 P_0 S$. Time $\tau = t/t_c$ is normalized by the characteristic time $t_c = \hbar/(2e I_c R_0)$. The derivation is given in the Supplemental Material~\cite{SM}.

\begin{table}[t]
    \centering
    \caption{Physical and dimensionless parameters used in simulations. Dimensionless values used for numerical simulations are rounded.}
    \label{tab:params}
    \begin{tabular}{ccc|cc}
        \hline\hline
        \multicolumn{3}{c|}{Physical} & \multicolumn{2}{c}{Dimensionless} \\
        \hline
        Symbol & Value & Unit & Symbol & Value \\
        \hline
        $d$ & 2 & \si{nm} & & \\
        $S$ & 0.25 & \si{\micro\meter\squared} & $\nu$ & 76 \\
        $I_c$ & 1 & \si{\micro\ampere} & & \\
        $C_0$ & 1 & \si{pF} & $\beta_c$ & 30 \\
        $R_0$ & 100 & \si{\ohm} & & \\
        $P_0$ & 0.1 & \si{\micro\coulomb\per\centi\meter\squared} & & \\
        $m_p$ & $1$\cite{morozovska2016} & $10^{-17}$ \si{\joule\meter\second\squared\per\coulomb\squared} & $\mu$ & 0.018 \\
        $\alpha$ & $1,2$\cite{scrymgeour2005,hlinka2006,rabe2007} & $10^8$ \si{\joule\meter\per\coulomb\squared} & $A$ & 2, 4 \\
        $\gamma_p$ & 4.77\cite{tang2022,tang2026a,fontana1994} & $10^{-5}$\si{\joule\meter\second\per\coulomb\squared} & $\gamma$ & 0.29 \\
        & & & $\tilde{\theta}$ & 0.2 \\
        \hline\hline
    \end{tabular}
\end{table}

Equations~\eqref{eq:ODEs-RCSJ}--\eqref{eq:ODEs-Ip} are solved numerically using Heun's method~\cite{SM,ruemelin1982,haller2023}. The initial conditions are $\mathcal{P}(0)=-1$, $\phi(0)=0$, $\mathcal{V}(0)=0$, and $\mathcal{I}_p(0)=0$. Table~\ref{tab:params} lists the physical and dimensionless parameters used in the simulations. These parameters correspond to a typical underdamped junction with plasma frequency $\omega_{JJ}=\sqrt{2eI_c/\hbar C_0}$ in the \si{\giga\hertz} range, while the ferroelectric mode $\omega_{FE}=\sqrt{2\alpha/m_p}$ lies in the \si{\tera\hertz} range. The chosen polarization $P_0=0.1~\mu$C/cm$^2$ is smaller than that of many conventional ferroelectrics such as perovskites and \ce{HfO2}-based ferroelectrics but is on par with 2D ferroelectrics like \ce{CuInP2S6} \cite{zhou2024,maisonneuve1997,liu2016,si2019,morozovska2024}, and keeps the linear approximation in Eq.~\eqref{eq:RCSJ-modified} valid and facilitates current-induced polarization switching, since a larger $P_0$ would deepen the double-well minima and suppress polarization switching.

\begin{figure}[ht]
    \centering
    \includegraphics[width=0.9\columnwidth]{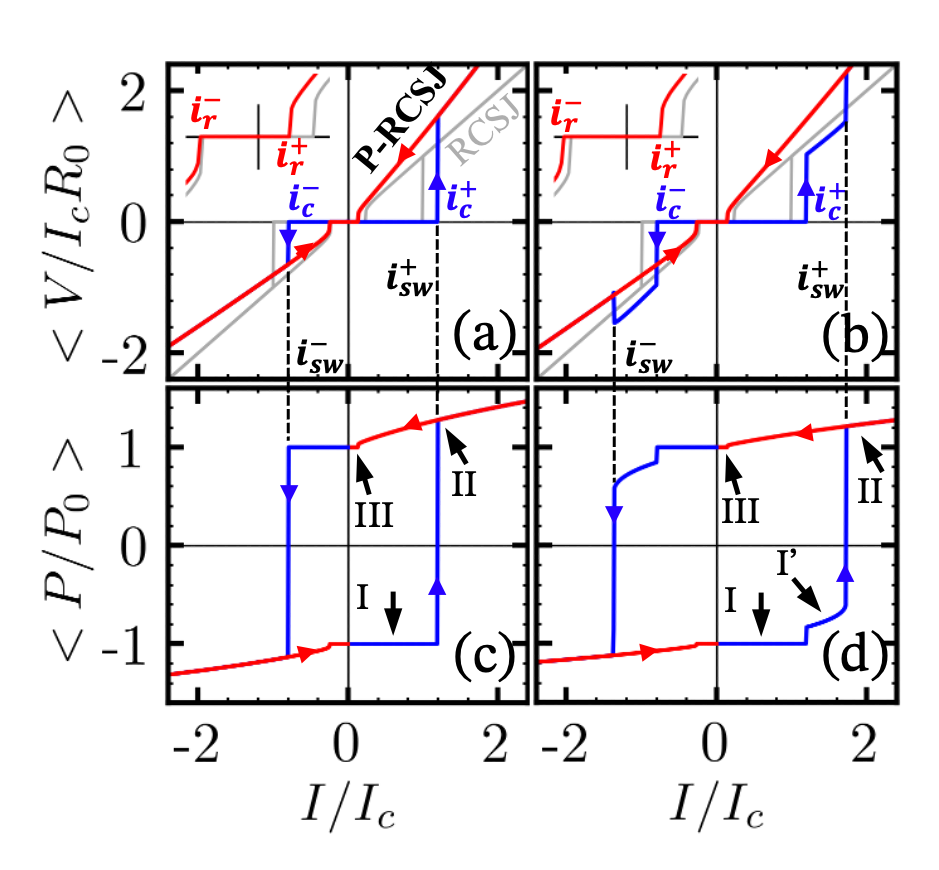}
    \caption{\textbf{Hysteresis in time-averaged voltage and polarization.} 
    (a), (b) Time-averaged voltage $ \mathcal{V}=V/I_cR_0 $ under current sweeps for Landau coefficients $A=2, 4$. 
    Blue/red curves are RCSJ results with electric polarization (P-RCSJ), while gray curves are conventional RCSJ results without ferroelectrics for comparison. 
    The blue and red curves show RCSJ results with electric polarization (P-RCSJ), while the gray curves show conventional RCSJ results without ferroelectricity for comparison.
    Insets: zoom-in near retrapping currents.
    (c), (d) Corresponding time-averaged polarization $\mathcal{P}=P/P_0 $ under current sweeps. The initial polarization is $\mathcal{P}(0)=-1$.}
    \label{fig:avg-plot}
\end{figure}


\textit{Nonreciprocity at zero temperature.---}For each bias current, we evolve the system to steady state and extract the time-averaged voltage $\mathcal{V}=V/I_cR_0$ and polarization $\mathcal{P}=P/P_0$. Figure~\ref{fig:avg-plot} shows the resulting I-V and I-P characteristics for two Landau coefficients, $A=2$ and $A=4$. The current is swept from zero to a positive maximum and back, followed by the corresponding negative sweep. The critical currents $i_c^\pm$ mark the transition from the superconducting to the resistive state, while the retrapping currents $i_r^\pm$ mark the return to the zero-voltage state. Compared with the conventional RCSJ model, shown by the gray curves, the FE-JJ exhibits asymmetric critical and retrapping currents, i.e., a pseudo-superconducting-diode effect.

This asymmetry follows directly from the polarization dependence in Eq.~\eqref{eq:ODEs-RCSJ} and is triggered by polarization switching, as shown in Fig.~\hyperref[fig:avg-plot]{2(c)}. Starting from $\mathcal{P}=-1$, the positive critical current is enhanced to $i_c^+=1+\tilde{\theta}$. Once the junction enters the resistive state, the finite voltage drives the polarization toward the opposite minimum, $\mathcal{P}=+1$, thereby producing a polarization switch. The subsequent negative sweep therefore probes this opposite polarization state, for which the critical-current magnitude is reduced to $|i_c^-|=1-\tilde{\theta}$. The diode efficiency, defined by $\eta = (|i_c^+|-|i_c^-|)/(|i_c^+|+|i_c^-|)$, is then equal to the asymmetry factor, $\eta = \tilde{\theta} = \theta P_0$, which is highly tunable through the junction design and the spontaneous polarization~\cite{tang2026}. A discussion of initial conditions is given in the Supplemental Material~\cite{SM}.

Thus, the nonreciprocity is not an equilibrium asymmetry of the current-phase relation but a history-dependent effect generated by polarization switching during the current sweep. This distinction is essential. If the critical current is independent of the polarization, i.e., $\tilde{\theta}=0$, or if the sweep amplitude remains below the polarization-switching threshold, the tunneling conductance remains unchanged throughout the sweep and the critical currents are identical; the pseudo-superconducting-diode effect therefore disappears.

We also note that, in an underdamped junction, the asymmetries in the critical and retrapping currents are not necessarily linked. For instance, the retrapping currents can differ even if polarization switching does not occur during the sweep. This is because the critical current is determined only by the polarization state at the moment of transition, whereas the retrapping current is also influenced by the history of the polarization dynamics in the resistive state. That history can differ because the coupled phase dynamics tilt the effective double-well potential asymmetrically near its minima.

Increasing the Landau coefficient $A$ deepens the ferroelectric double-well potential and suppresses polarization switching by increasing the required voltage. For $A=4$, the system first enters an intermediate resistive state with fractional average polarization $|\mathcal{P}|<1$ [Fig.~\ref{fig:avg-plot}(b,d)]. In this state, the superconducting phase evolves while the polarization remains trapped near its original minimum. Further increasing the bias destabilizes this metastable state and triggers full polarization switching. According to Eq.~\eqref{eq:RCSJ-modified}, the two resistive states have different resistances: $R_0/(1+\tilde{\theta})$ and $R_0/(1-\tilde{\theta})$.

\begin{figure}[ht]
    \centering
    \includegraphics[width=1\columnwidth]{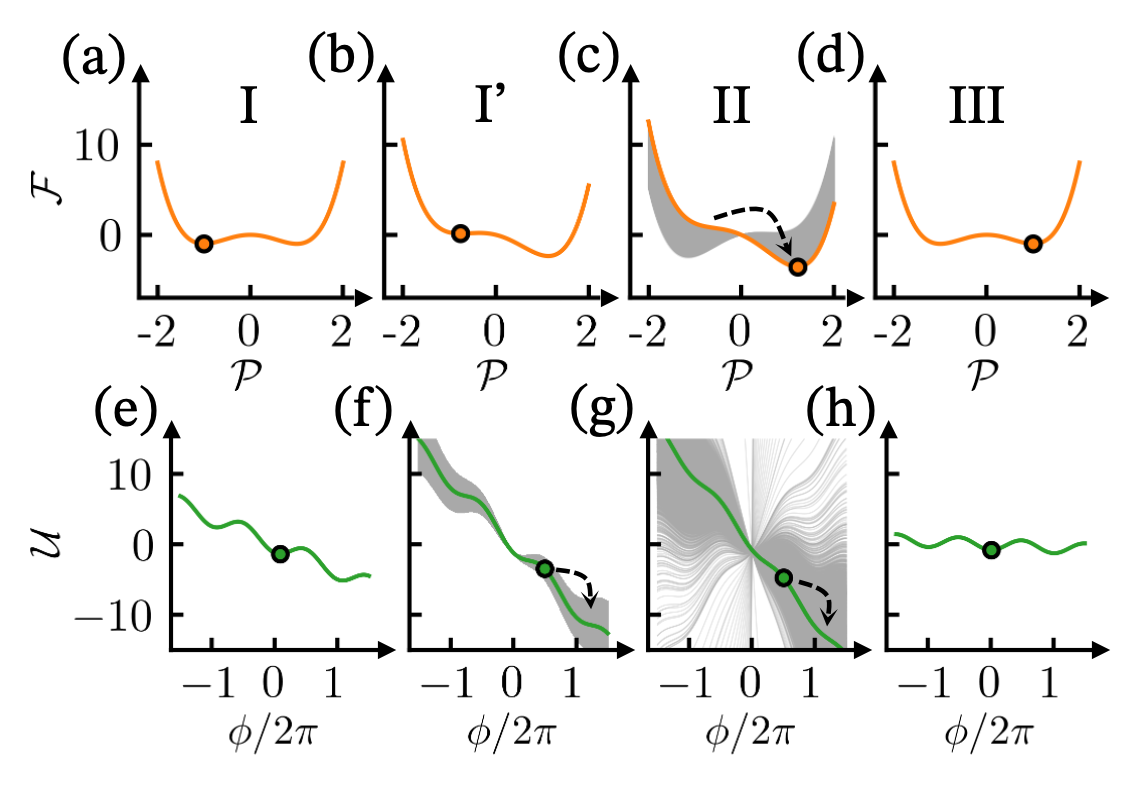}
    \caption{\textbf{Rocking double-well and washboard potentials.} 
    Potential landscapes are constructed using the time-averaged data and correspond, from left to right, to sweep points \Romannum{1}, \Romannum{1}', \Romannum{2}, and \Romannum{3} in Fig.~\hyperref[fig:avg-plot]{2(d)}.
    Gray regions/curves indicate oscillation amplitudes. 
    (a)--(d) Ferroelectric double-well potential profiles (orange curves) and polarization values (orange balls).
    (e)--(h) Corresponding Josephson-junction washboard potentials (green curves) and phase values (green balls).
    }
    \label{fig:potentials}
\end{figure}

Figure~\ref{fig:potentials} illustrates the double-well and washboard potentials for different states along the intermediate branch for $A=4$ in Fig.~\hyperref[fig:avg-plot]{2(b,d)}. After the phase escapes, the finite voltage shifts the double-well potential and drives polarization oscillations, producing a polarization current that rocks the washboard potential. Because the double-well potential is deep (i.e., the barrier is high), the polarization remains trapped near its original minimum, giving rise to a metastable resistive state [Fig.~\hyperref[fig:potentials]{3(b,f)}]. At larger bias, this metastable state breaks down: the double-well barrier is suppressed by the voltage, and the polarization begins to switch [Fig.~\hyperref[fig:potentials]{3(c,g)}].


\textit{Thermal effects.---}Because the pseudo-superconducting-diode effect is dynamical and relies on polarization switching, it is important to assess its robustness against thermal fluctuations. We include thermal noise in both the Josephson and ferroelectric dynamics, leading to the following stochastic differential equations (SDEs):

\begin{align}
     \mathcal{I} &= \beta_c \dot{\mathcal{V}} + (1-\tilde{\theta}\mathcal{P})\mathcal{V} + (1-\tilde{\theta} \mathcal{P})\sin\phi + \mathcal{I}_p + \sigma_{JJ}\Gamma(\tau), \label{eq:ODEs-RCSJ-thermal} \\
    \mathcal{V} &= \mu \ddot{\mathcal{P}} + \gamma \dot{\mathcal{P}} + A(\mathcal{P}^2-1)\mathcal{P} + \frac{\tilde{\theta}}{\nu}\cos\phi + \sigma_{FE}\Gamma'(\tau), \label{eq:ODEs-LKT-thermal}
\end{align}
where $\Gamma(\tau)$ and $\Gamma'(\tau)$ are independent Gaussian white noises with zero mean and unit correlations. We have ignored the termperature dependence of $I_c$ for simplicity. The fluctuation-dissipation theorem gives the noise strengths as \cite{SM}

\begin{equation}
    \sigma_{JJ} = \sqrt{\frac{2k_BT}{I_c\hbar/2e}},  \  
    \sigma_{FE} = \sqrt{\frac{2k_BT}{I_c\hbar/2e} \frac{\gamma}{\nu}}, \label{eq:sigma}
\end{equation}
with $k_B$ the Boltzmann constant and $T$ the temperature.

\begin{figure}[ht]
    \centering
    \includegraphics[width=0.9\columnwidth]{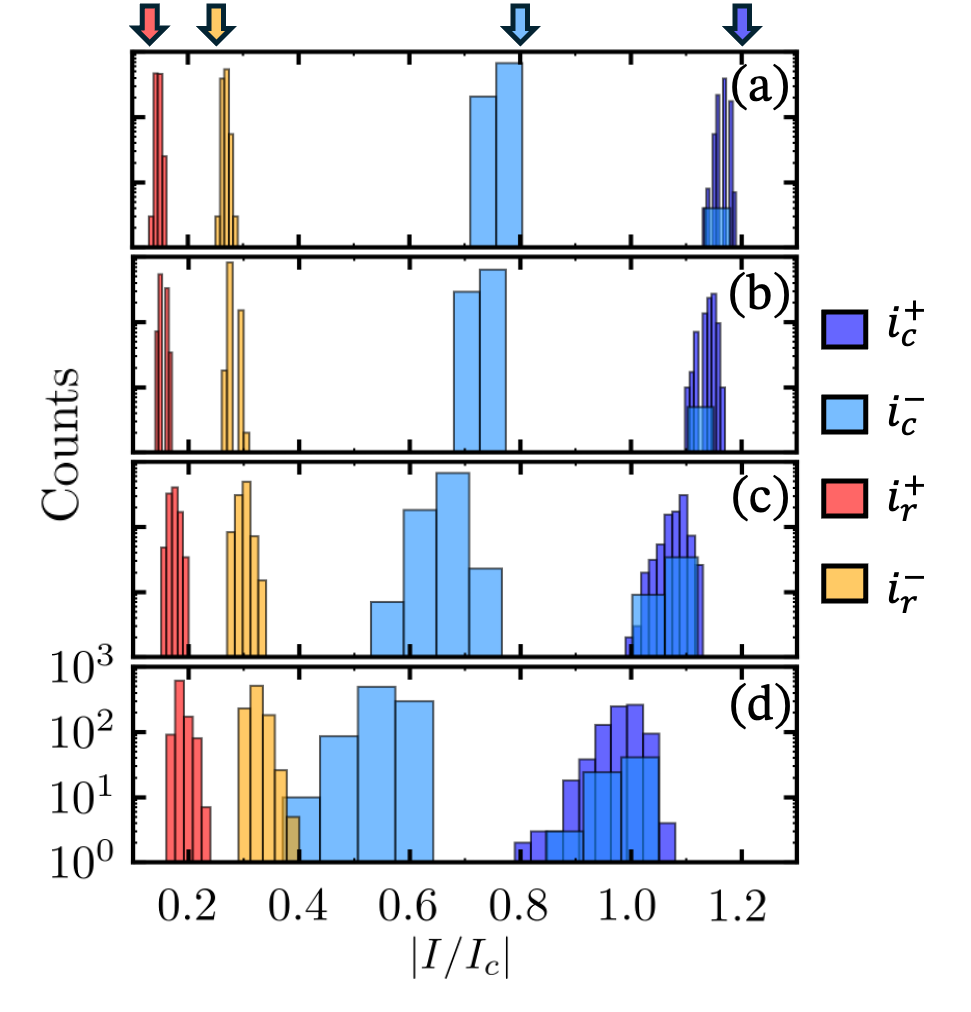}
    \caption{\textbf{Histograms of critical ($i_c^{\pm}$) and retrapping ($i_r^{\pm}$) currents.} The temperature $T$ is set to (a) 0.1 K, (b) 0.2 K, (c) 0.5 K, and (d) 1 K. Each histogram (shown on a logarithmic scale) is constructed from 1000 full current sweeps. The arrows at the top indicate the zero-temperature values of the critical and retrapping currents for reference. The critical temperature of the superconductor is estimated to be $T_c \simeq 0.5$ K using simple BCS approximation.}
    \label{fig:histograms}
\end{figure}

We perform 1000 current sweeps at each temperature, $T=0.1$, $0.2$, $0.5$, and $1.0$ K, and extract the critical and retrapping currents from each I--V curve. The resulting histograms are shown in Fig.~\ref{fig:histograms}. The asymmetries between $|i_c^+|$ and $|i_c^-|$, as well as between $|i_r^+|$ and $|i_r^-|$, persist at finite temperature. Increasing $T$ broadens the distributions, shifts the critical currents to lower values, and shifts the retrapping currents to higher values, consistent with thermally assisted switching and retrapping.

Notably, a distinct secondary component of $i_c^-$ emerges at finite temperature, as indicated by the overlap between the $|i_c^+|$ and $|i_c^-|$ distributions in Fig.~\ref{fig:histograms}. For the parameters used here, the ferroelectric barrier energy is much larger than the thermal energy, so this component is not caused by thermally activated polarization reversal. Instead, it likely reflects more complex phase--polarization dynamics induced by thermal fluctuations in the resistive state.

\begin{figure}[ht]
    \centering
    \includegraphics[width=1\columnwidth]{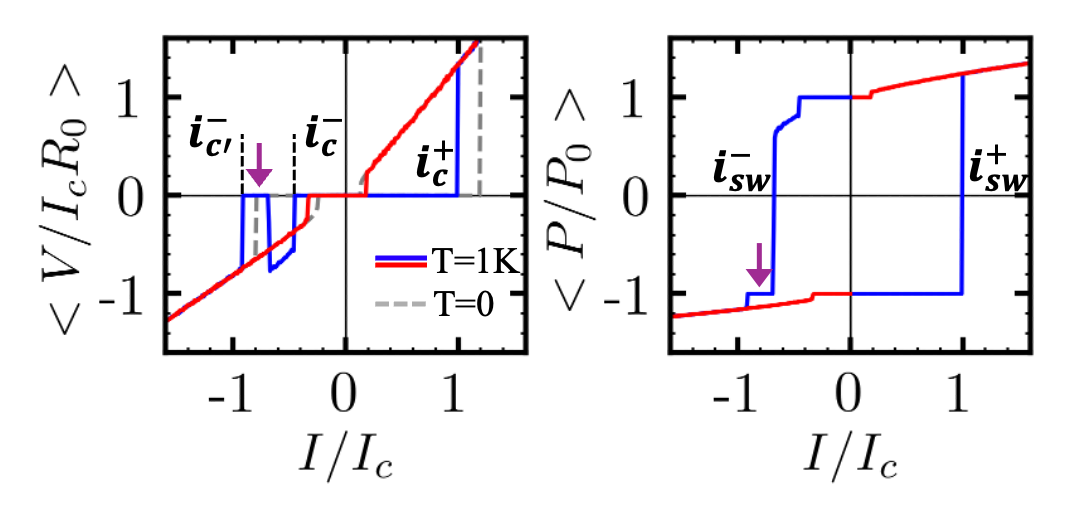}
    \caption{\textbf{Time-averaged characteristics at $T=1$ K.} Time-averaged (a) voltage and (b) polarization under current sweeps. The purple arrow indicates the additional superconducting region. The sweep protocol and initial conditions are the same as in Fig.~\hyperref[fig:avg-plot]{2(a)}. Dashed gray curves in (a) show the zero-temperature results for comparison.}
    \label{fig:avg-plot-thermal}
\end{figure}

Figure~\ref{fig:avg-plot-thermal} shows a representative current sweep at $T=1$ K. Compared with the zero-temperature result, shown by the dashed gray curves, thermal fluctuations induce premature switching into the resistive state and earlier retrapping. In the negative branch, an additional superconducting region appears between two resistive transitions. The system first enters the resistive state by thermal activation at $i_c^-$, then retraps into the superconducting state when the polarization switches at $i_{sw}^-$, before transitioning again to the resistive state at $i^-_{c'}$. 

The reappearance of the superconducting state results from the combined effects of thermal fluctuations and polarization switching. Due to premature thermal activation, the phase initially moves through the washboard potential at a lower rate, corresponding to a small voltage. When polarization switching occurs at $i^-_{sw}$, the critical-current magnitude increases from $1-\tilde{\theta}$ to $1+\tilde{\theta}$, thereby raising the Josephson energy barrier. As a result, the phase particle does not have sufficient kinetic energy to overcome the barrier and retraps into the superconducting state. Further increasing the bias current eventually drives a second transition to the resistive state at $i^-_{c'}$.


\textit{Conclusions.---}In summary, we have proposed a pseudo-superconducting-diode effect in inversion-asymmetric ferroelectric Josephson junctions. By coupling a polarization-dependent RCSJ model to Landau-Khalatnikov-Tani ferroelectric dynamics, we showed that current-induced polarization switching produces unequal critical and retrapping currents for opposite sweep directions. The effect is intrinsically nonequilibrium: it depends on the current-sweep history and disappears when the polarization does not switch. Consequently, materials with relatively small polarization, such as \ce{CuInP2S6}~\cite{zhou2024,maisonneuve1997,liu2016,si2019,morozovska2024}, and ultrathin ferroelectric layers are particularly favorable. The resulting nonreciprocity is tunable through ferroelectric parameters and the sweep protocol and remains robust against thermal fluctuations. Our results identify FE-JJs as electrically programmable platforms for magnetic-field-free nonreciprocal superconducting devices.


\textit{Acknowledgments.---}The authors would also like to thank the Kavli Foundation for its support through the Kavli Institute Innovation Award and the Kavli Institute of Nanoscience Delft. M.N.A. acknowledges support of the NWO Talent Programme VIDI financed by the NWO VI.Vidi.223.089.

\bibliographystyle{apsrev4-2}
\bibliography{main}


\clearpage

\onecolumngrid

\appendix

\begin{center}
{\large \textbf{Supplemental Material for}}\\[0.5em]
{\large \textbf{\emph{"Pseudo-superconducting-diode effect in ferroelectric Josephson junctions"}}}
\end{center}

\renewcommand{\theequation}{S\arabic{equation}}
\renewcommand{\thefigure}{S\arabic{figure}}
\renewcommand{\thetable}{S\arabic{table}}
\renewcommand{\thepage}{S\arabic{page}}
\setcounter{equation}{0}
\setcounter{figure}{0}
\setcounter{table}{0}
\setcounter{page}{1}


\section*{Derivation of the FE-RCSJ model via Lagrangian formalism}
\label{sec:appendix-model}

The kinetic energy of the system consists of two contributions: The capacitive energy of the Josephson junction and the kinetic energy of the ferroelectric polarization:

\begin{align}
    T_{JJ} &= \frac{C_0}{2}(\frac{\hbar}{2e}\dot{\phi})^2, \\
    T_{FE} &= \int_{vol} \frac{m_p}{2} \dot{P}^2 dV,
\end{align}
where the volume integral is taken over the ferroelectric barrier. The potential energy is given by the sum of the Josephson energy and the ferroelectric free energy:

\begin{align}
    U_{JJ} &= \frac{\hbar}{2e} \left[ -\left(1-\theta P \right)I_c\cos\phi - I\phi \right], \\
    U_{FE} &= \int_{vol} \left[ -\frac{\alpha}{2}P^2 + \frac{\alpha}{4}\frac{P^4}{P_0^2} - \left(\frac{1}{d}\frac{\hbar}{2e}\dot{\phi}P \right) \right] dV.
\end{align}
The modulation of Josephson energy by the polarization is included via the $(1-\theta P)$ factor. The first two terms in $U_{FE}$ correspond to the Landau free energy, while the last term accounts for the interaction between the electric field and the polarization. The Lagrangian of the system is then given by $\mathcal{L} = T_{JJ} + T_{FE} - U_{JJ} - U_{FE}$. For simplicity, we replace the volume integral by the volume $Sd$, where $S$ is the junction area and $d$ the barrier thickness.

Applying the Lagrangian formalism to the variables $\phi$ and $P$,

\begin{align}
    \frac{d}{dt}\frac{\partial\mathcal{L}}{\partial \dot{\phi}} - \frac{\partial\mathcal{L}}{\partial \phi} &= 0, \\
    \frac{d}{dt}\frac{\partial\mathcal{L}}{\partial \dot{P}} - \frac{\partial\mathcal{L}}{\partial P} &= 0,
\end{align}
and including damping/dissipation terms, we obtain the coupled equations of motion:

\begin{align}
    0 &= \left(\frac{\hbar}{2e} \right)^2C_0\ddot{\phi} + \frac{\hbar}{2e}\left(1-\theta P_0 \frac{P}{P_0}\right)\frac{1}{R_0} \left(\frac{\hbar}{2e}\Dot{\phi} \right) + \frac{\hbar}{2e}\left[ \left(1-\theta P_0 \frac{P}{P_0} \right)I_c\sin\phi - I + S\Dot{P} \right], \label{eq:appendix-RCSJ} \\
    0 &= dm_p\Ddot{P} + d\gamma_p \Dot{P} + d\alpha \left(\frac{P^2}{P_0^2}-1 \right)P + \frac{\hbar}{2e}\frac{I_c}{SP_0}\theta P_0 \cos \phi - \frac{\hbar}{2e}\Dot{\phi}. \label{eq:appendix-FE}
\end{align}
The polarization dependence of the resistive current in Eq.~\eqref{eq:appendix-RCSJ} arises because the tunneling conductance is modulated by the polarization \cite{tang2026}. The amplitude of the voltage-shift term $\sim \cos\phi$ in Eq.~\eqref{eq:appendix-FE} is proportional to the asymmetry factor $\theta$ and the Josephson energy. This term shifts the ferroelectric double-well potential through the superconducting phase, even in the zero-voltage state. Consequently, the polarization minima are slightly displaced from $\mathcal{P}=\pm 1$. For realistic parameters, this displacement is negligible and does not affect our main results, as confirmed numerically. We therefore retain the term in the simulations, while neglecting its effect in the qualitative discussion for simplicity. Normalizing the equations using the dimensionless variables and parameters defined in the main text, we arrive at Eqs.~\eqref{eq:ODEs-RCSJ}--\eqref{eq:ODEs-Ip}.

\section*{Influence of initial conditions}
\label{sec:appendix-initial-conds}

\begin{figure}[ht]
    \centering
    \includegraphics[width=0.5\columnwidth]{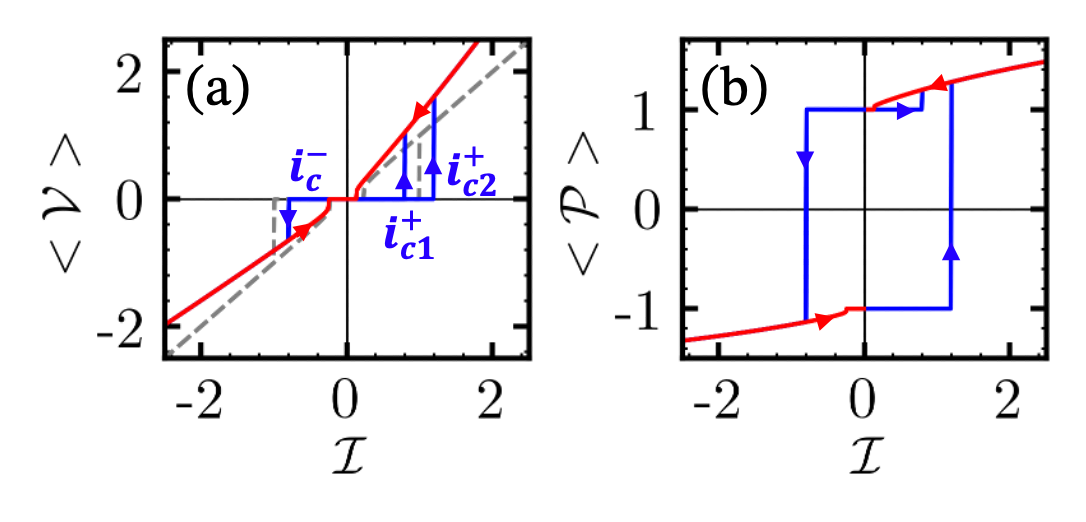}
    \caption{\textbf{Time-averaged voltage and polarization.} Initial polarization is set at $\mathcal{P}(0)=+1$, Landau coefficient $A=2$, and the current sweep is looped twice.}
    \label{fig:supp-2loops}
\end{figure}

Figure~\ref{fig:supp-2loops} shows the time-averaged voltage and polarization for an initial polarization $\mathcal{P}(0)=+1$ when the bias current goes two times through the loop. The Landau coefficient is set to $A=2$, while the other parameters are kept the same as in Table~\ref{tab:params}. In the first loop (preparatory sweep), the critical currents become symmetric ($i^+_{c1} = i^-_c$), so the pseudo-superconducting-diode effect vanishes for the critical currents, while an asymmetry persists in the retrapping currents. This follows from the polarization dynamics in Fig.~\hyperref[fig:supp-2loops]{S1(b)}: the superconducting state has $\mathcal{P}=+1$ for both sweep directions in the first loop, yielding identical critical currents $i_{c,1}^+ = i_{c,1}^- = 1-\Tilde{\theta}$, while in the resistive state the polarization switches to the negative minimum, producing different retrapping currents. Because the polarization remains negative at the start of the second loop, the critical current in that loop shifts ($i^+_{c1} = 1+\Tilde{\theta}$) and the pseudo-superconducting-diode behavior reappears in the subsequent IV and IP traces.

\section*{Time evolution}
\label{sec:appendix-time-evolution}

\begin{figure}[ht]
    \centering
    \includegraphics[width=0.6\columnwidth]{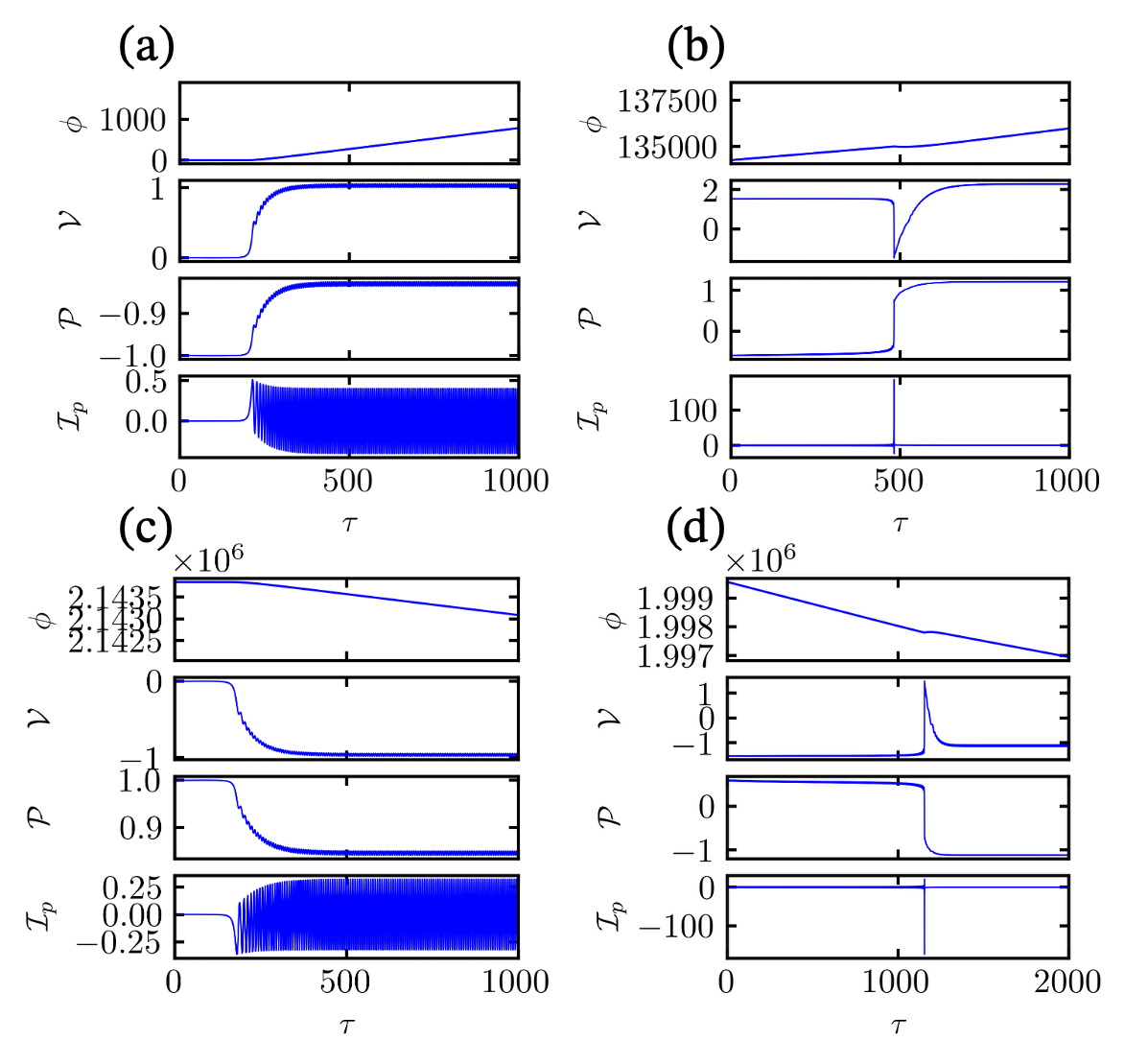}
    \caption{\textbf{Time evolution.} Time evolution of phase $\phi$, voltage $\mathcal{V}$, polarization $\mathcal{P}$, and polarization current $\mathcal{I}_p$. Landau coefficient $A=4$. Panels (a) and (b) correspond to the metastable \Romannum{1}' and switching state \Romannum{2} in Fig.~\hyperref[fig:avg-plot]{2(d)}, respectively; (c) and (d) correspond to the same states in the negative sweeping branch. The time scale is chosen for better visibility.}
    \label{fig:supp-time}
\end{figure}

For deeper double-well potentials, the polarization can be trapped in a metastable state before switching, as shown in Fig.~\hyperref[fig:avg-plot]{2(d)} and Fig.~\hyperref[fig:potentials]{3(b)}. The time evolution of the variables is plotted in Fig.~\hyperref[fig:supp-time]{S2(a,c)}. When the system enters the metastable state, a finite voltage appears and the phase starts to wind accordingly. The polarization starts to oscillate and finally settles at a fractional value $|\mathcal{P}|<1$. The polarization current, given by Eq.~\eqref{eq:ODEs-Ip}, fluctuates around zero with moderate amplitude, weakly rocking the washboard potential.

Figure~\hyperref[fig:supp-time]{S2(b,d)} show the time evolution during polarization switching. The polarization undergoes a rapid change from one minimum to the other, accompanied by a large transient polarization current. The sudden increase/decrease in polarization current suppresses/enhances the displacement and normal resistive currents, leading to a temporary drop/spike in the voltage. Such transient voltage changes feed back on the polarization dynamics, hindering further growth of the polarization current until the polarization has settled.

\section*{Resonance features}
\label{sec:appendix-resonance}

\begin{figure}[ht]
    \centering
    \includegraphics[width=0.5\columnwidth]{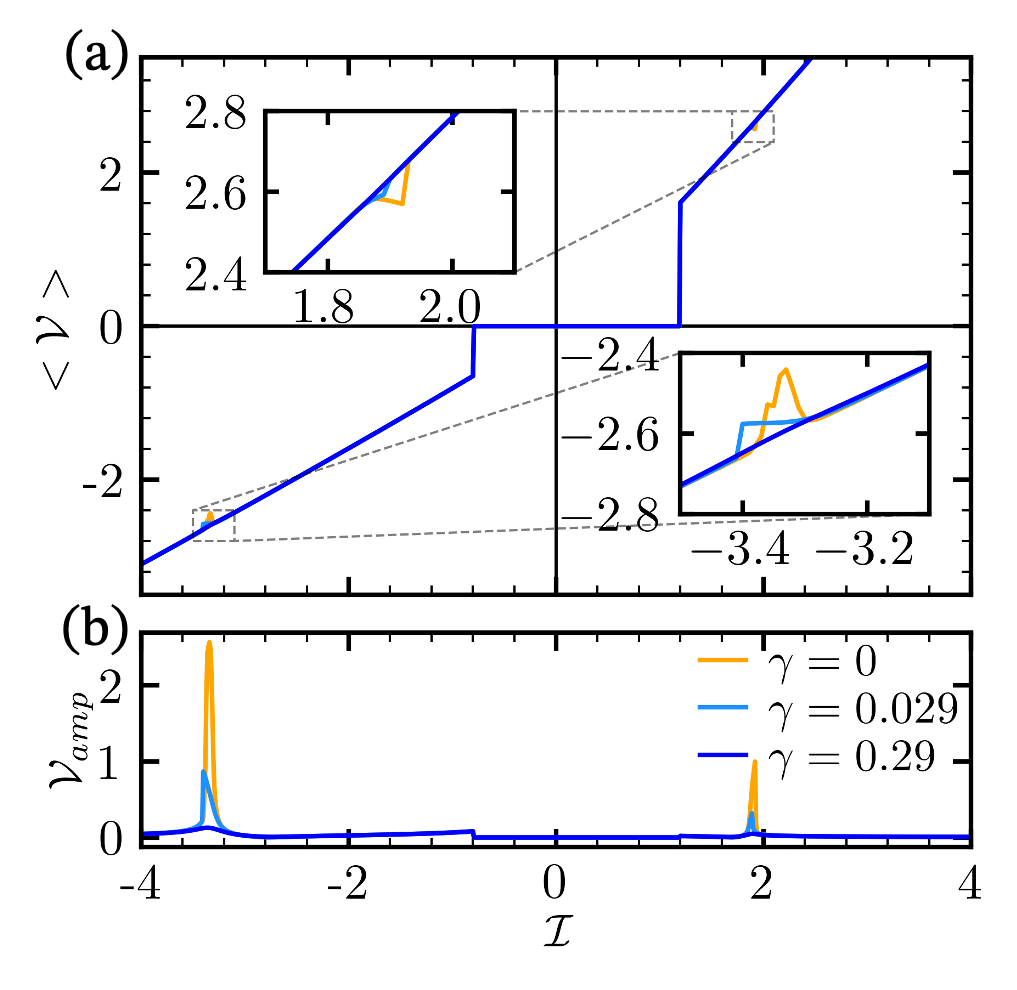}
    \caption{\textbf{Time-averaged voltage and voltage amplitudes.} Time-averaged (a) voltage and (b) voltage amplitude under current sweeps for different damping constants $\gamma$. }
    \label{fig:supp-resonance}
\end{figure}

In the main text, we discuss the more experimentally relevant case where the characteristic frequency of the ferroelectric is much larger (around 2 orders of magnitude) than the Josephson plasma frequency. Here, in order to explore additional features of the coupled dynamics, we consider a larger polarization inertia $m_p=10^{-15}$ \si{\joule\meter\second\squared\per\coulomb\squared}, which reduces the ferroelectric frequency to the Gagahertz range ($\sim 71$ \si{\giga\hertz} for $\alpha=10^8$ \si{\joule\meter\per\coulomb\squared}). The other parameters are the same as in Table~\ref{tab:params}. Figure~\hyperref[fig:supp-resonance]{S3(a)} shows that, in addition to the pseudo-superconducting-diode effect, step-like features appear in the IV characteristics at reasonable voltage values. Figure~\hyperref[fig:supp-resonance]{S3(b)} confirms that these steps correspond to resonances between the Josephson and ferroelectric modes. The resonance features become more pronounced for smaller damping $\gamma$, as expected \cite{rahmonov:hal-03295760}.


\section*{Stochastic Differential Equations and Numerical Method}
\label{sec:appendix-SDE}

To include thermal fluctuations in both the Josephson and ferroelectric dynamics, we add fluctuating noise terms to Eqs.~\eqref{eq:appendix-RCSJ} and \eqref{eq:appendix-FE}, yielding the following stochastic differential equations (SDEs):

\begin{align}
    \frac{\hbar}{2e}C_0\ddot{\phi} + \left(1-\theta P_0 \frac{P}{P_0}\right)\frac{1}{R_0}\left(\frac{\hbar}{2e}\Dot{\phi}\right) + \left(1-\theta P_0 \frac{P}{P_0} \right)I_c\sin\phi + S\Dot{P} + \delta I_{JJ} &= I, \label{eq:appendix-SDE-RCSJ} \\
    \frac{dm_p}{S}(S\Ddot{P}) + \frac{d\gamma_p}{S} (S\Dot{P}) + \frac{d\alpha}{S} \left(\frac{P^2}{P_0^2}-1 \right)SP + \frac{\hbar}{2e}\frac{I_c}{SP_0}\theta P_0 \cos \phi + \delta V_{FE} &= \frac{\hbar}{2e}\Dot{\phi}, \label{eq:appendix-SDE-FE}
\end{align}
where $SP$ is the polarization charge, $dm_p/S$ has the dimension of inductance, $d\gamma_p/S$ of resistance, and $d\alpha/S$ of inverse capacitance. The fluctuating current $\delta I_{JJ}$ and voltage $\delta V_{FE}$ are white noises with zero mean and correlation functions:

\begin{align}
    \langle \delta I_{JJ} (t) \delta I_{JJ} (t') \rangle &= \frac{2k_BT}{R_0}\delta(t-t'), \\
    \langle \delta V_{FE} (t) \delta V_{FE} (t') \rangle &= 2k_BT \frac{d\gamma_p}{S}\delta(t-t'),
\end{align}
where $\delta(t)$ is the Dirac delta function. Introducing the normalized time $\tau = t/t_c$ with characteristic time $t_c = \hbar/2eI_cR_0$, the correlation functions satisfy

\begin{align}
    \langle \delta I_{JJ} (t) \delta I_{JJ} (t') \rangle &= \frac{1}{t_c} \langle \delta I_{JJ} (\tau) \delta I_{JJ} (\tau') \rangle, \\
    \langle \delta V_{FE} (t) \delta V_{FE} (t') \rangle &=  \frac{1}{t_c} \langle \delta V_{FE} (\tau) \delta V_{FE} (\tau') \rangle.
\end{align}
Therefore, the correlation functions in normalized units read

\begin{align}
    \langle \delta \mathcal{I}_{JJ} (\tau) \delta \mathcal{I}_{JJ} (\tau') \rangle &= \sigma_{JJ} ^2 \delta(\tau-\tau'), \\
    \langle \delta \mathcal{V}_{FE} (\tau) \delta \mathcal{V}_{FE} (\tau') \rangle &= \sigma_{FE} ^2 \delta(\tau-\tau'),
\end{align}
with the noise amplitudes given by Eq.~\eqref{eq:sigma} in the main text.

The SDEs are solved numerically using Heun's method \cite{ruemelin1982,haller2023}. By introducing the variable $\textbf{x} = \left[ \phi, \dot{\phi}, P, \dot{P} \right]$, we convert the two second-order differential equations, Eqs.~\eqref{eq:ODEs-RCSJ-thermal} and \eqref{eq:ODEs-LKT-thermal}, into four first-order equations:

\begin{align}
    \frac{d}{d\tau} x_1 &= x_2, \\
    \frac{d}{d\tau} x_2 &= \frac{1}{\beta_c} \left[ \mathcal{I} - (1-\Tilde{\theta} x_3)x_2 - (1-\Tilde{\theta} x_3)\sin{(x_1)} - \nu x_4 + \sigma_{JJ}\frac{N(0,d\tau)}{d\tau} \right], \\
    \frac{d}{d\tau} x_3 &= x_4, \\
    \frac{d}{d\tau} x_4 &= \frac{1}{\mu} \left[ x_2 - \gamma x_4 - A(x_3^2-1)x_3 - \frac{\Tilde{\theta}}{\nu}\cos{(x_1)} + \sigma_{FE}\frac{N'(0,d\tau)}{d\tau} \right],
\end{align}
where we use the definition of Gaussian white noise: $\Gamma(\tau) = N(0,d\tau)/d\tau$, with $N(0,d\tau)$ a normal distribution with zero mean and variance $d\tau$.

Heun's method is a two-step updating scheme. We calculate the variable $\textbf{x}_{n+1}$ at time step $n+1$ from $\textbf{x}_{n}$ at time step $n$ as follows. In the first step, we calculate the Euler predictor:

\begin{equation}
    \textbf{x}_{n+1} = \textbf{x}_{n} + h\textbf{K}_{n} + \textbf{F},
\end{equation}
where $h$ is the time step size and $\textbf{K}_{n}$ is the gradient vector evaluated at time step $n$, given by

\begin{equation}
    \textbf{K}_n = 
    \begin{bmatrix}
        x_{2,n} \\
        \frac{1}{\beta_c} \left[ \mathcal{I} - (1-\Tilde{\theta} x_{3,n})x_{2,n} - (1-\Tilde{\theta} x_{3,n})\sin{(x_{1,n})} - \nu x_{4,n} \right] \\
        x_{4,n} \\
        \frac{1}{\mu} \left[ x_{2,n} - \gamma x_{4,n} - A(x_{3,n}^2-1)x_{3,n} - \frac{\Tilde{\theta}}{\nu}\cos{(x_{1,n})} \right]
    \end{bmatrix},
\end{equation}
and $\textbf{F}$ is the thermal noise contribution,

\begin{equation}
    \textbf{F} = 
    \begin{bmatrix}
        0 \\
        \frac{\sigma_{JJ}}{\beta_c} N(0,h) \\
        0 \\
        \frac{\sigma_{FE}}{\mu} N'(0,h)
    \end{bmatrix},
\end{equation}
where $N(0,h)$ and $N'(0,h)$ are two random numbers drawn from normal distributions with zero mean and variance $h$.

In the second step, we calculate the corrected gradient vector using the average $(\textbf{K}_n + \textbf{K}_{n+1})/2$ evaluated with the predictor $\textbf{x}_{n+1}$, and the final update is given by

\begin{equation}
    \textbf{x'}_{n+1} = \textbf{x}_{n} + \frac{h}{2}(\textbf{K}_n + \textbf{K}_{n+1}) + \textbf{F}.
\end{equation}

For non-stochastic cases at zero temperature, we set $\sigma_{JJ}=\sigma_{FE}=0$, and the same procedure applies.

\section*{I-V characteristics at finite temperatures}
\label{sec:appendix-thermal}

\begin{figure}[ht]
    \centering
    \includegraphics[width=0.5\columnwidth]{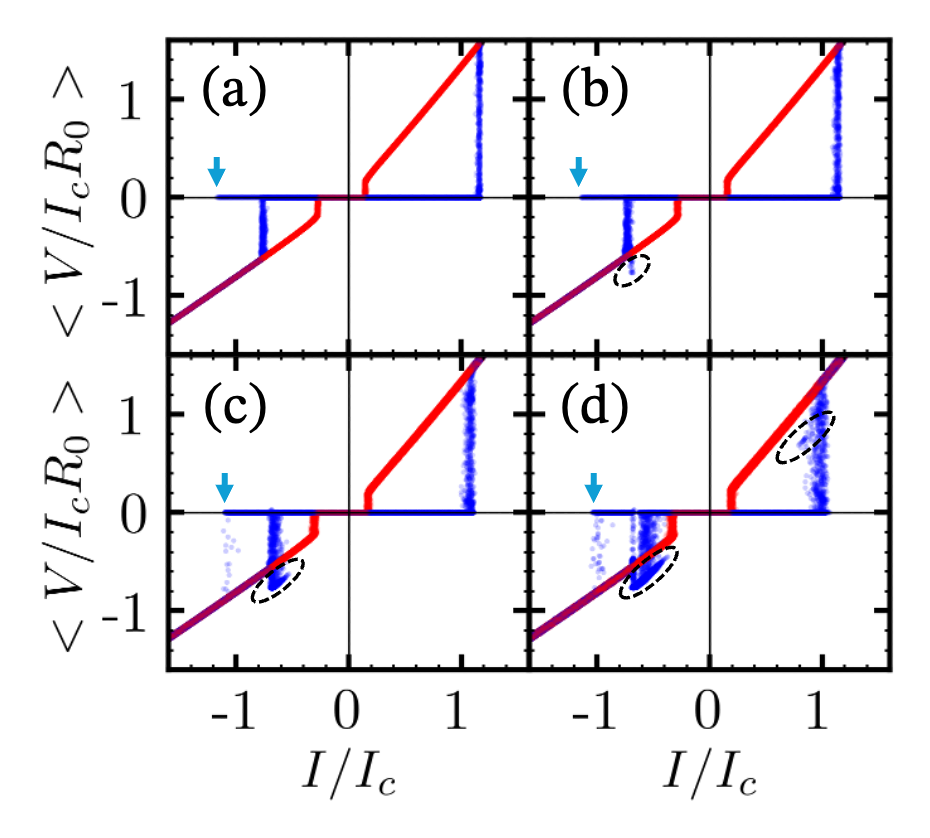}
    \caption{\textbf{I-V scatter plots at finite temperatures.} The tempature $T$ is set to (a) 0.1 K, (b) 0.2 K, (c) 0.5 K, and (d) 1 K. Each scatter plot is constructed from 1000 full current sweeps. The sweeping protocol and initial conditions are the same as in Fig.~\hyperref[fig:avg-plot]{2(a)}. The dashed black circles highlight the intermediate resistive states induced by thermal noise, and the light blue arrows indicate the symmetric components of the critical currents (see text).}
    \label{fig:supp-avg-scatter-thermal}
\end{figure}

Figure~\hyperref[fig:supp-avg-scatter-thermal]{S4} shows the scatter plots of the IV characteristics at different finite temperatures. Comparing with the zero-temperature case in Fig.~\hyperref[fig:avg-plot]{2(a)}, we find that thermal noise also gives rise to intermediate resistive states, highlighted by the dashed black circles. These states are similar to the metastable resistive states observed for larger Landau coefficient $A$ in Fig.~\hyperref[fig:avg-plot]{2(b)}. This behavior occurs because thermal fluctuations can drive the superconducting phase into a resistive state at lower bias currents. The resulting small voltage, however, is insufficient to switch the polarization until the bias current, and hence the voltage, is increased further.

We further find that the intermediate resistive states are more likely to appear on the negative sweeping branch at lower temperatures. This asymmetry arises because the initial polarization is negative, which makes the Josephson energy barrier higher on the positive sweeping branch and therefore suppresses thermal activation there, requiring a higher temperature for the intermediate resistive state to appear.

\end{document}